\documentstyle[preprint,aps]{revtex}
\begin{document}
\draft
\title{
Quasifree photoabsorption on neutron-proton pairs in
$^3$He\thanks{Supported by the Deutsche Forschungsgemeinschaft
(SFB 201) and the Academy of Finland}}
\author{J. A. Niskanen$^{1)}$, P. Wilhelm$^{2)}$, and
H. Arenh\"ovel$^{2)}$}
\address{
1) Department of Theoretical Physics, Box 9, FIN-00014 University
of Helsinki, Finland \\
2) Institut f\"ur Kernphysik,
Johannes Gutenberg-Universit\"at, D-55099 Mainz, Germany}
\date{\today}
\maketitle
\begin{abstract}
Three-body photodisintegration of $^3$He is calculated in the photon
energy range 200 -- 400 MeV assuming quasifree absorption on
$np$ pairs both in initial quasideuteron and singlet
configurations.  The model includes the normal
nucleonic current, explicit meson exchange currents
and the $\Delta(1232)$-isobar excitation.  The total cross
section is increased by a factor of about 1.5 compared with free
deuteron photodisintegration. Well below and above the
$\Delta$ region also some spin observables differ
significantly from the ones of
free deuteron disintegration due to the more compressed wave
function of the correlated $np$ pairs in $^3$He compared to
the deuteron.
The initial singlet state causes a significant change in the
analyzing power $A_y$. These differences
could presumably be seen at the conjugate angles where two-body
effects are maximized and where photoreactions could complement
similar pion absorption experiments.
\end{abstract}
\newpage
\narrowtext
\section{Introduction}

Recently, there has been
considerable experimental activity in studies of pion
absorption on two or three nucleons in light nuclei \cite{weyer}.
These can be expected to illuminate
the role of strong dynamics in few-nucleon systems and also
modifications of pion absorption due to the nuclear environment.
Certainly, the most direct
comparison can be made between absorption on
the deuteron and on correlated nucleon pairs in
quasifree kinematics. The latter can be arranged by choosing
the so called conjugate angles, which maximise the role of the
two-nucleon mechanism, so that the reaction closely resembles
the absorption on a deuteron
\cite{aniol,altman,weber,mukho}.

Parallel to this experimental activity concentrating on cross section
measurements, model predictions were published
not only for cross sections but also
for spin observables both in positive \cite{pihe} and
negative \cite{piminus} pion absorption on nucleon pairs in $^3$He.
The former reaction suggested sensitivity of polarization observables
on the quasideuteron wave function, which is not the case
for the angular distribution. In the denser $^3$He-nucleus
short-ranged rescattering mechanisms are weighted differently
than the long-ranged but oscillatory direct nucleon-nucleon
overlap as compared with the free deuteron reaction. However,
at least the experimental outgoing proton polarization, although
measured for the quasifree kinematics, does not agree with the
prediction from the quasifree model \cite{may}. The origin of
this discrepancy is not known presently, but one possibility
would be the strong initial state interaction of the pion with the
target nucleus. In the future, more data on quasifree absorption
of positive pions are expected for both the outgoing proton
polarization and also for the analyzing power $iT_{11}$ of a
polarized
$^3$He target \cite{e445}, which might help to analyze these
discrepancies in greater detail.

Contrary to the disagreement in the positive pion case, a
quantitative agreement is obtained for negative pion absorption
between data on a proton pair \cite{aniol,altman,weber,mukho} and
theoretical results \cite{piminus,heavy}. It is interesting to note
that in this case a new short-range meson exchange contribution to
the two-nucleon axial current, suggested by Lee and Riska \cite{lee},
has a significant effect \cite{heavy}. Of course, a bound diproton
does not exist as such, so a direct  comparison between free and
quasifree process is not possible as in the case for the absorption
on a quasideuteron. However, if in the inverse process $ n + p
\rightarrow p + p + \pi^-$ the outcoming protons have a small
relative momentum, then at this low energy the resonant $^1S_0$
scattering state may be considered as nearly bound and can be used
with some theoretical input for a comparison with the two-body
absorption in $^3$He. So far, the actual experiments have been
performed as $\vec{p} + n \rightarrow (p + p)_{^1S_0} + \pi^-$ but on
a deuteron target \cite{ponting,e460} and the spin observables show
little deviation from the predictions. However, in this case there is
only a rather moderate dependence on the relative two-body wave
function. Still, the accuracy of the experiments is sufficient in
order to allow in principle to distinguish between several wave
functions.

With the above somewhat confused situation in pion absorption --
agreement between theory and experiment in one case, strong
discrepancy in the other -- one is tempted to consider another probe
for the study of the pair correlations. If the source of the
disagreement is, indeed, the strong initial state $\pi$--nucleus
interaction, then an obvious choice would be to attempt a similar
approach with an electromagnetic probe. In this case the initial
state interaction between the probe and the target nucleus is
negligible and one may study the quasi two-nucleon processes in a
much cleaner way. Another bonus is that with different quantum
numbers due to different couplings this reaction offers somewhat
complementary information with respect to pion absorption.

The aim of this paper is to study explicitly to which extent
observables in medium energy photodisintegration of $^3$He in
quasifree two-body situations depend on the initial pair wave
function. To a large extent this study will be devoted to the
dependence of spin variables which were predicted to be sensitive in
the case of pion absorption. It may be suited  for experimental
tests, e.g., at the LEGS facility at Brookhaven with polarized
photons \cite{legs}. Preliminary results for cross sections have been
published in Ref.\ \cite{physlet}.

There is a large amount of work on photodisintegration of both the
deuteron \cite{laget,anas,leide,tanabe,wilhelm} and $^3$He
\cite{lagethe} also at intermediate energies. The model of
\cite{leide,wilhelm} to be used in this paper has described deuteron
photodisintegration rather successfully in the $\Delta
(1232)$-resonance region. It is now applied to the quasifree
situation with modified pair wave functions. The differences between
the predictions using modified pair wave functions and those for the
free deuteron case should give an idea about the trends how the
experimental results may differ from the free reaction.  Conversely,
if the theoretical results are sensitive enough to the details of the
assumed pair wave functions, the experimental differences may reveal
properties of those. This approach may also allow one to estimate the
inportance of three-body effects once the two-body absorption is
understood.

Earlier theoretical work on photodisintegration of $^3$He
\cite{lagethe} has considered various one-, two- and three-nucleon
mechanisms in the excitation function of one outcoming nucleon at
some particular angle. Also most experimental results have been
presented in this way \cite{dhose,ruth}. The first results with
polarized photons on $^3$He$(\vec\gamma,p)X$ from LEGS confirm the
prediction by Laget about the importance of three-nucleon effects in
general on the one hand and about the existence of a quasi two-body
region on the other hand, where two-body mechanisms are by far
dominant \cite{ruth}. However, this earlier work does not concentrate
on any detailed investigations in the quasifree region, which should
be possible at the right momentum and at conjugate angles appropriate
to two-body absorption and will be a further subject of this paper as
a complement of two-body pion absorption studies.

The only published experimental results close to our explicit
two-nucleon approach are those of the kinematically complete
tagged-photon experiment of the TAGX collaboration \cite{tagx}. There
the differential cross sections of protons and neutrons were
presented. Most of the neutron cross section can be considered as
arising from the absorption on a quasideuteron in $^3$He, and even in
the proton cross section there are clear indications of a separation
into active fast protons and slow spectators.

In addition to the absence of any initial state interaction, there is
another significant difference in photodisintegration compared to
positive pion absorption. In pion absorption the existence of a
neutron-proton pair in the isovector $^1S_0$ state has a negligible
effect because of two suppressions of important mechanisms
\cite{mix}. Firstly, due to the conservation of parity and angular
momentum, the final states are spin triplets with $J \neq L$ and the
$N\Delta$ admixture can never be in an $S$-state as in the absorption
on a quasideuteron. Further, $s$-wave pion rescattering is restricted
to the weak isospin symmetric pion-nucleon amplitude in the nucleonic
isospin conserving transition $^1S_0 \rightarrow\, ^{3}\! P_0$. In
the photon case a similar suppression may be true for the
$\Delta$-dominated M1 transitions. But according to the
Thomas-Reiche-Kuhn sum rule E1 transitions should be as important for
the singlet as for the triplet pair. The only suppression is a
statistical factor of 3 since there are 3/2 $np$-pairs in the triplet
state and only 1/2 in the singlet. So, outside the $\Delta$ region
also this initial state could be significant.

In the next section we discuss some details of the model, in
particular the nucleon sector with a quasideuteron pair embedded in
the wave function of $^3$He. The results are presented in Section 3
and Section 4 gives a summary and conclusions.

\section{Model}

\subsection{Nucleon sector}

The simple starting point for the $^3$He ground state is to describe
its spin-isospin structure with total spin projection $m$ by the
totally antisymmetric state \cite{piminus}
\begin{equation}
|\Psi_m\rangle
= \frac{1}{\sqrt{2}} \left( \left[ \left[ \frac{1}{2} \times
\frac{1}{2} \right]^{01} \times \frac{1}{2}
\right]^{\frac{1}{2}\frac{1}{2}, \frac{1}{2}m} - \left[
\left[ \frac{1}{2} \times \frac{1}{2} \right]^{10} \times
\frac{1}{2}\right]^{\frac{1}{2}
\frac{1}{2}, \frac{1}{2}m} \right) |\Psi_{space}\rangle
\label{eq:1}
\end{equation}
assuming a symmetric $S$-wave space part. Using Jacobi coordinates
\begin{equation}
\vec{r}_{ij}=\vec{r}_i-\vec{r}_j,\quad\quad
\vec{\rho}_k=\frac{1}{2}\, (\vec{r}_i+\vec{r}_j)-\vec{r}_k\quad
(ijk\mbox{ cyclic}),
\end{equation}
where $\vec{r}_i$ are the individual particle coordinates, it is
parametrized in the form
\begin{equation}
\Psi_{space}(\vec{r}_1,\vec{r}_2,\vec{r}_3)
= \frac{v(r_{12})}{r_{12}}\,\frac{u(\rho_3)}{\rho_3}
 + (\mbox{cyclic permutations}).
\end{equation}
The square bracket in Eq.\ (\ref{eq:1}) denotes the usual
coupling to a good total spin and isospin state
\begin{equation}
\left[\frac{1}{2} \times \frac{1}{2} \right]^{TT_z, SS_z} =
\sum_{\sigma_1\sigma_2\tau_1\tau_2}
\langle
\frac{1}{2}\sigma_1\frac{1}{2}\sigma_2 | SS_z \rangle
\,\langle
\frac{1}{2}\tau_1\frac{1}{2}\tau_2 | TT_z \rangle\,
| \frac{1}{2} \sigma_1, \frac{1}{2} \tau_1 \rangle\,
| \frac{1}{2} \sigma_2, \frac{1}{2} \tau_2 \rangle\, .
\end{equation}
The ``quasideuteron'' can be identified in the first term of Eq.\
(\ref{eq:1}). By writing the outermost coupling explicitly, one can
exhibit separately the spin-isospin structures of the spectator and
the pair. In this form, assuming the spectator to be inactive in the
reaction, its role is diminished to just carrying a known amount of
spin and charge without otherwise affecting the two-body reaction
amplitudes. The total wave function now can be written as
\begin{equation}
|\Psi_m\rangle={\cal A}\,|\psi_m\rangle
\label{eq:5}
\end{equation}
with
\begin{eqnarray}
|\psi_{\pm\frac{1}{2}}\rangle &=&
\pm\sqrt{\frac{1}{3}}\, |d,\pm 1\rangle\, |u(p),\mp\frac{1}{2}\rangle
\mp\sqrt{\frac{1}{6}}\, |d,0    \rangle\, |u(p),\pm\frac{1}{2}\rangle
\nonumber\\ & &
+\sqrt{\frac{1}{6}}\, |s(pn)    \rangle\, |u(p),\pm\frac{1}{2}\rangle
-\sqrt{\frac{1}{3}}\, |s(pp)    \rangle\, |u(n),\pm\frac{1}{2}\rangle\,
 . \label{eq:pp}
\end{eqnarray}
being antisymmetric with respect to $1\leftrightarrow 2$ only and
\begin{equation}
{\cal A}=\frac{1}{\sqrt{3}}\, (1-P_{13}-P_{23})
\end{equation}
to achieve a total antisymmetrization where $P_{ij}$ denotes
interchange of particles $i$ and $j$. The quasideuteron wave function
with the magnetic quantum number $\mu$ has been denoted by
$|d,\mu\rangle$, with $|s\rangle$ correspondingly the $^1S_0$
isovector pair with its proton and neutron contents explicitly shown
and with $|u(p/n),m_{sp}\rangle$ the spectator proton/neutron wave
function with spin projection $m_{sp}$. Explicitly the pair wave
functions read in the coordinate representation
\begin{eqnarray}
\langle \vec r\, | d,\mu\rangle & = &\frac{ v_d(r)}{r}\,
Y_{00}(\hat r)\,
\left[\frac{1}{2} \times \frac{1}{2} \right]^{00,1\mu} \nonumber\\
& & + \frac{ w_d(r)}{r} \sum_{M_L M_S} \langle 2M_L 1M_S |
1\mu\rangle\,
Y_{2M_L}(\hat r)\,\left[\frac{1}{2} \times \frac{1}{2}
\right]^{00,1M_S}  \nonumber \\
\langle \vec r\, | s(pn) \rangle & = & \frac{v_s (r)}{r}\,
Y_{00}(\hat r)\,
\left[ \frac{1}{2} \times \frac{1}{2} \right]^{10,00} \;.
\label{eq:tb}
\end{eqnarray}
In the form of Eq.\ (\ref{eq:5}), the space parts of the different
pair wave functions have now no longer be assumed to be identical
$S$-waves. In particular, the quasideuteron can be generalized to
include also the $D$-state in its relative correlation as shown in
Eq.\ (\ref{eq:tb}).  The
active pair in the last term of Eq.\ (\ref{eq:pp}) consists of two
protons and is not of concern in the present context modelling the
disintegration of $^3$He in terms of a quasideuteron. However, the
third term is in principle indistinguishable from the quasideuteron
processes and has to be considered.

Next, we want to specify the two-nucleon wave functions which have
been used. A product wave function, symmetric in all the three
coordinates, is a possible way to describe the spatial structure of
$^3$He. Then the relative two-nucleon wave function could reasonably
be taken as the square root of a correlation function for the pair
density. This description produces the static properties quite well
\cite{hadji}. As correlation function we shall use the isoscalar one
calculated from the Faddeev equations using the Reid soft core
potential and given by Friar et al. \cite{friar}. It is supplemented
by the $D$-state component ($P_D=10.5\%$) in the case of the $T = 0$
quasideuteron as explained in \cite{pihe}. This wave function is
significantly compressed towards shorter distances as compared with
the free deuteron one for which we use the one of the Bonn OBEPR
potential \cite{bonn}. Another possible choice for the spatial part
of $\Psi$ is a three-term parametrization in terms of basic states of
the two Jacobi coordinates in different permutations given by Hajduk
et al. \cite{hajduk}
\begin{equation}
\Psi_{LM\,lm} (\vec{r}_1,\vec{r}_2,\vec{r}_3) =
\frac{v_L(r_{12})}{r_{12}} \,\frac{ u_l(\rho_3)}{\rho_3} \,
Y_{LM}(\hat r_{12})\, Y_{lm}(\hat \rho_3) +
(\mbox{cyclic permutations}).
\end{equation}
In the present work, however, we need only the terms where the active
nucleon pair is in the two-nucleon relative wave function $v$. The
integrals of the other permutations may be thought to be simulated in
the correlation function treatment. At least this is the case for the
static properties. The tiny component of the spectator $D$-wave is
also omitted here. Since integration over the spectator degrees of
freedom is implied in any case, this omission is completely
insignificant. Fig.\ 1 shows a comparison of the spatial wave
functions used in this work.

Having in mind the photon absorption on a neutron proton pair, an
appropriate final state basis is given by
\begin{equation}
|\Psi_{SM,m_{sp}}\rangle={\cal A}\,
\left(|\vec{P}_{np},\phi_{SM}\rangle\, |\vec{p}_{sp}m_{sp}\rangle\right)\, .
\label{eq:final}
\end{equation}
The spectator proton is described by a plane wave with momentum
$\vec{p}_{sp}$ and spin projection $m_{sp}$. The scattering wave
function of the outgoing two fast nucleons with total momentum
$\vec{P}_{np}$ is denoted by $|\vec{P}_{np},\phi_{SM}\rangle$ with
spin $S$ and projection $M$. Of course, the states of Eq.\
(\ref{eq:final}) can be assumed to be approximately orthogonal only
in the restricted kinematic region we are interested in and which is
characterized by a slow proton and a fast $np$ pair. To calculate
$|\phi_{SM}\rangle$ the Lippmann-Schwinger equation for the Bonn
OBEPR potential \cite{bonn} has been solved. In the isospin one
states, $N\Delta$ components are taken into account by means of a
coupled $NN$-$N\Delta$ calculation in momentum space which allows a
good reproduction of the $NN$ scattering phase shifts. In particular,
the phase shift in the $^1D_2$ partial wave is well described
\cite{wilhelm}. This channel is of crucial importance in the $\Delta$
region because of its coupling to the $^5S_2(N\Delta)$ partial wave
with vanishing angular momentum barrier. Its magnetic dipole
excitation clearly dominates deuteron photodisintegration in the
resonance region.

The energy of the two-nucleon final state wave function is obtained
by two-body kinematic relations considering the pair to be bound by
10$\,$MeV more than the deuteron. One half of this shift is due to
the actual binding energy difference, the other half allows an
average of 5$\,$MeV kinetic energy for the spectator. It may be noted
that the square of the momentum space wave function of the spectator
of Ref. \cite{hajduk}, weighted by the square of the momentum, is
peaked at this energy. Also the spectator momentum distribution
observed in Ref. \cite{tagx} is peaked around 100$\,$MeV$/c$
corresponding to about 5$\,$MeV kinetic energy. This simple
description was found to simulate the exact two-nucleon energy very
well for pion absorption in Ref.~\cite{piminus}.

\subsection{Amplitudes and observables}

The $8\times4$ transition matrix for three-body photodisintegration
of the $^3$He nucleus reads \begin{equation}
\delta(\vec{P}_{np}+\vec{p}_{sp}-\vec{k})\, {\cal
M}_{m_{sp},SM,\lambda,m}=-\left\langle\Psi_{SM,m_{sp}}|\,
\vec{\epsilon}_{\lambda}\cdot\vec{J}(\vec{k})\, |\Psi_m\right\rangle,
\label{eq:amp} \end{equation} where $\vec{\epsilon}_{\lambda}$ is the
photon polarization vector, $\vec{k}$ the photon momentum and
$\vec{J}$ the electromagnetic current which includes one-body and
two-body parts \begin{equation}
\vec{J}=\sum_i\vec{J}_i+\sum_{i<j}\vec{J}_{ij}\, . \end{equation}
Some more details will be given in the next section. As it stands,
Eq.\ (\ref{eq:amp}) of course also contains non-diagonal matrix
elements, where the two-body current involves a nucleon which does
not belong to a correlated initial or final pair. However, they are
assumed to be negligible because of the short-range nature of the
two-body mechanism and since we have to resrict ourself to reaction
kinematics where a slow proton and a fast $np$ pair are observed.
Thus the amplitude can finally be expressed in terms of the pure
two-body amplitudes which read
\begin{equation}
{\cal M}^{d}_{SM,\lambda,\mu}=
-\left\langle\phi_{SM}|\,
\vec{\epsilon}_{\lambda}\cdot\vec{J}\, |d,\mu\right\rangle,
\quad\quad\quad
{\cal M}^{s}_{SM,\lambda}=
-\left\langle\phi_{SM}|\,
\vec{\epsilon}_{\lambda}\cdot\vec{J}\, |s\right\rangle,
\end{equation}
for deuteron and $^1S_0(np)$ photodisintegration, respectively.
One finds
\begin{equation}
{\cal M}_{m_{sp},SM,\lambda,\pm\frac{1}{2}}
=\left[
\pm\sqrt{\frac{1}{3}}\, \delta_{m_{sp},\mp\frac{1}{2}}
{\cal M}^{d}_{SM,\lambda,\pm 1}
+\sqrt{\frac{1}{6}}\, \delta_{m_{sp},\pm\frac{1}{2}}
\left( {\cal M}^{s}_{SM,\lambda} \mp {\cal M}^{d}_{SM,\lambda,0}
\right) \right] \tilde{u}(-\vec{p}_{sp}).
\label{eq:11}
\end{equation}
The momentum space wave function $\tilde{u}$, reflecting the relative
Fermi motion of the spectator with respect to the initial pair, will
disappear after integration over the spectator momentum. Moreover,
since the polarization of the spectator is not observed, the final
density matrix is diagonal with respect to $m_{sp}$, i.e.,
$\tau_f=\tau\,\delta_{m_{sp}'m_{sp}}$, where $\tau$ describes the
density matrix of the pair spin degrees. In case of an unpolarized
$^3$He target, Eq.\ (\ref{eq:11}) leads to
\begin{equation}
\int d^3p_{sp}\,\mbox{tr} ({\cal M}^\dagger \tau {\cal M} \tau_i) =
 \frac{1}{6}\, \mbox{tr} ({\cal M}^{d\dagger} \tau {\cal M}^d \tau_{\gamma})
+\frac{1}{6}\, \mbox{tr} ({\cal M}^{s\dagger} \tau {\cal M}^s \tau_{\gamma}),
\label{eq:tau}
\end{equation}
where $\tau_\gamma$ denotes the initial photon density matrix. This
means that one ends up with an incoherent sum of the quasideuteron
and the $^1S_0$ pair contributions. Expressed in terms of the
deuteron and the singlet disintegration cross sections for
unpolarized photons which are
\begin{equation}
\frac{d \sigma_d}{d\Omega_p}
= \frac{1}{6}\, \mbox{tr}({\cal M}^{d\dagger} {\cal M}^d),\quad\quad
\frac{d \sigma_s}{d\Omega_p}
= \frac{1}{2}\, \mbox{tr}({\cal M}^{s\dagger} {\cal M}^s),
\label{eq:sd}
\end{equation}
respectively, the $^3$He disintegration cross section reads
\begin{equation}
\frac{d \sigma_{\mbox{\scriptsize He}}}{d\Omega_p} = \frac{1}{2}\,
 \frac{d \sigma_d}{d\Omega_p}
+ \frac{1}{6}\, \frac{d \sigma_s}{d\Omega_p}\, .
\label{eq:he}
\end{equation}
Here it is important to note that the cross section in Eq.\
(\ref{eq:he}) refers to an active particle coordinate, say the active
proton. It should not be mixed up with the one-arm proton cross
section of the reaction $^{3}$He$(\gamma,p)pn$ measurement. Actually
it is not a directly measurable quantity, but has to be extracted
from a kinematically complete experiment after an assumption on the
spectator momentum distribution has been made. Also, analogously to
the total cross section \cite{gw}, a statistical factor 1/3 has been
inserted in Eq.\ (\ref{eq:he}) to account for the
indistinguishability of the unobserved spectator.

For linearly polarized photons, Eq.\ (\ref{eq:tau}) leads to the
following relation for the photon asymmetry
\begin{equation}
\Sigma_{\mbox{\scriptsize He}}=
\frac{\Sigma_d+\frac{\displaystyle 1}{\displaystyle 3}\,
\frac{\displaystyle d\sigma_s}{\displaystyle d\sigma_d}\,\Sigma_s}
{1+\frac{\displaystyle 1}{\displaystyle 3}\,
\frac{\displaystyle d\sigma_s}{\displaystyle d\sigma_d}} \, .
\label{eq:sig}
\end{equation}
Again, the photon asymmetry  refers to photon polarization parallel
and perpendicular to the plane which is defined by the photon and the
active proton momenta. A completely analogous expression to Eq.\
(\ref{eq:sig}) is obtained for the polarization of the outgoing fast
nucleons. Thus for unpolarized $^3$He the contributions from the
triplet and singlet initial pairs are decoupled. If, say, the $^1S_0$
pair contribution is much smaller than the quasideuteron
contribution, it can be omitted in first order approximation. This
would mean that the polarizations of the two fast nucleons for
example are then the same for both reactions $\gamma +
^3$He$\rightarrow (pn) + p_{sp}$ and $\gamma + d \rightarrow p + n$.

Considering transversely polarized $^3$He the analyzing power $A_y$
is given by
\begin{eqnarray}
A_y\, \frac{d\sigma_{\rm He}}{d\Omega_p}
& = & \int d^3p_{sp}\,\frac{1}{2}\,
\mbox{tr} ({\cal M}^\dagger {\cal M} \sigma_y) \\ \nonumber
& = & \frac{\sqrt{2}}{3} \,\mbox{Im}\, \sum_{SM} \left[
 \left({\cal M}^{d}_{SM,0,+1}\right)^{\star}
 \left({\cal M}^{d}_{SM,+1,+1}-{\cal M}^{d}_{SM,-1,+1}\right)\right.
 \\ \nonumber
&   &\left. \quad\quad \quad\quad\quad
+\left({\cal M}^{s}_{SM,   +1}\right)^{\star}
 \left({\cal M}^{d}_{SM,+1,+1}+{\cal M}^{d}_{SM,-1,+1}\right)
 \right].
\end{eqnarray}
This is the same result as given in \cite{mix} for pion absorption on
$^3$He and is also the same as for photon (or pion) absorption on a
free deuteron, except for the presence of the $^1S_0$ pair as well as
different wave functions and slightly different kinematics. Finally,
it may be worth noting the reason for the preference of the Cartesian
component $A_y$ over the spherical one $it_{11}$ of the analyzing
power in comparisons of the $^3$He and deuteron reactions. In the
Madison convention \cite{madison} the $A_y$ has the same expression
in both cases in terms of the transition matrices, so that any
differences would have a basically dynamic origin. However, the
spherical quantities for the spin-$\frac{1}{2}$ and spin-1 particles
would have normalizations differing by an additional factor of
$\sqrt{2/3}$.

\subsection{Electromagnetic interaction}

The photoabsorption mechanisms on the two-nucleon system which have
been included in the present calculation are summarized in Fig.\ 2.
The model includes the usual one-nucleon current (N[1]) ( Fig.\ 2a),
which is given by the spin and the convection current. Furthermore,
the spin orbit current which gives the most important relativistic
contribution is also considered. Moreover, Siegert operators
corresponding to the nonrelativistic one-body charge density are
applied. Their use allows to take into account the dominant part of
the exchange current contribution to the electric multipoles in a
model independent way (see e.g. \cite{arenhrev}). The electromagnetic
interaction described up to now defines the so called normal part
(N).

In case of the quasideuteron disintegration explicit static $\pi$- as
well as $\rho$-meson exchange currents (MEC) beyond the Siegert
operators are included. In the nucleonic sector shown in Fig.\ 2b,c,
they are consistent with respect to gauge invariance to the $\pi$-
and to the dominant part of $\rho$-exchange in the OBEPR potential as
explained in detail in Ref.~\cite{schmitt}. In the calculation of the
$^1S_0(np)$ disintegration the MEC effects are incorporated via the
Siegert operators only.

Of course, direct $\Delta$ excitation is the most important
photoabsorption mechanism at intermediate energies. Within the
$NN$-$N\Delta$ coupled channel approach it can be considered as a
one-body contribution depicted in Fig.\ 2d, once the $N\Delta$
component of the wave function has been generated. We take into
account the dominant magnetic dipole excitation only using the
modified $\gamma N\Delta$ coupling of Ref.~\cite{wilhelm} which led
to a good description of the size and the energy dependence of the
total cross section for deuteron photodisintegration in the $\Delta$
region. The two-body $\Delta$ excitation due to the exchange current
shown in Fig.\ 2e-g is of minor importance.

It is worth noting that all the $\Delta$-excitation mechanisms of
Fig.\ 2d-g cannot contribute to the break-up of the $^1S_0(np)$ pair
because of the isospin selection rule. Since the $\Delta$ excitation
is always going along with an isovector transition, it cannot link
neutron-proton isovector states having $T_z=0$, i.e., the
contributions from neutron and proton excitation cancel each other
exactly. This argument is not valid any more, if the break-up of the
$^1S_0(pp)$ pair would be considered.

\section{Results}

The primary aim of this paper is to study the expected deviations of
the quasifree from the free two-nucleon disintegration due to
differences in the initial state wave functions as shown in Fig.\ 1.
However, it is of interest at first to make some comments about the
effect of the different mechanisms shown in Fig.\ 2 and discussed in
Sec.\ 2.3. Fig.\ 3 presents the accumulation of these contributions
at 300 MeV photon energy, just below the $\Delta$-resonance energy
which corresponds to $E_\gamma \approx 320$ MeV. All observables are
presented as a function of the proton angle in the center of momentum
system of the two fast final state nucleons (or of the photon and the
initial nucleon pair), which is the only free variable in two-body
reactions. These results are obtained using the square root of the
correlation function for the initial pair wave function. The dotted
curves show the purely one-nucleon current contribution and
short-dashed ones the one-body part with the inclusion of two-body
terms by the Siegert operators . In the long-dashed curves the normal
part is further supplemented with the explicit $\pi$- and
$\rho$-meson exchange current effects (MEC) shown in Fig.\ 2b,c
beyond the Siegert parts. The most important individual effect is the
direct $\Delta$ excitation shown in Fig.\ 2d, which is included in
the solid curves. Furthermore, the $\Delta$-MEC of Fig.\ 2e-g is
added in the solid curves, but it is of minor importance.

The dominance of the $\Delta$ isobar has been demonstrated earlier in
deuteron photodisintegration \cite{leide,tanabe,wilhelm}, but in the
present quasideuteron case the initial state wave function is
significantly more compressed to shorter distances. Thus one could
expect the short-range effects to be more prominent. Also the
nucleonic term can obtain stronger high momentum components with the
compressed wave function. So for the more condensed initial wave
function both the one-nucleon and MEC as well as the isobar effects
should be enhanced.

To investigate their interplay quantitatively and also for
completeness, Fig.\ 4 presents the same observables for the true
deuteron wave function and kinematics.  An enhancement by a factor of
three in the cross section can be attributed to the more condensed
wave function. Both the explicit MEC and the isobar effects on the
one hand and the nucleon current contribution on the other hand are
increased, but in the latter the enhancement is stronger. Here all
changes go in the same direction to increase the total cross section,
but the angular distribution remains the same at this energy. The
statistical factor $1\over 2$ in Eq.\ (\ref{eq:he}) reduces this
enhancement leaving a ratio of approximately 1.5 between the cross
sections of Figs. 3 and 4. Experimental evidence \cite{dhose,tagx}
indicates the ratio of this magnitude. It should be emphasized that
the difference of the initial wave functions is the reason for the
quasifree cross section being larger than the free one, {\it not} the
number of quasideuteron pairs in $^3$He.

In the photon asymmetry  $\Sigma$, the free deuteron gets relatively
larger individual contributions than $^3$He. In both cases the
explicit MEC and the isobar contribution go oppositely to the normal
part resulting in virtually indistinguishable total results. The
proton polarization is not changed significantly either in this
energy region with the maximal $\Delta$ contribution. All results for
the neutron polarization were similar to those for the proton and
thus will not be shown separately. Only the analyzing power $A_y$ has
a significantly different result at this energy, partly because of
the first order interference effect from the singlet state pair wave
function as shown in Eq.\ (\ref{eq:he}).

A detailed study of the contributions from the different components
of the initial pair wave functions is presented in Fig.\ 5. All cross
sections are normalized as contributions to
$d\sigma_{\mbox{\scriptsize He}}/d\Omega_p$ in Eq.\ (\ref{eq:he}).
The short-dashed curves show the results for the quasideuteron
component alone. Large parts of the amplitudes arise from its
$D$-state. Its omission leads to qualitative changes in the nucleonic
spin observables as seen in the long-dashed curves. This can be
expected because the nucleon spin orientation is totally different in
the $D$-state than in the $S$-state. The separate $^1S_0$
contribution (also in this case the spin observables are scaled with
the corresponding cross section) is also qualitatively different from
the quasideuteron, but its weight is overall too small to cause any
significant changes in the total observables except in $A_y$ (solid
vs.\ short-dashed curves). In the reaction products this initial
configuration is, of course, indistinguishable. However, with
different multipoles (notably E2) it can be seen separately for
example in the quasifree disintegration of the diproton in $^3$He
\cite{future}. It may be noted that $A_y$ is not existing in case of
the $^1S_0(np)$ disintegration.

Fig.\ 6 shows the dependence on the different models for the pair
wave functions at four photon energies ranging from 220 MeV to 360
MeV. The dotted curves are the results for a free deuteron pair wave
function corresponding to deuteron photodisintegration with the cross
section multiplied by the factor 1/2 of Eq.\ (\ref{eq:he}) for
comparison of the pure wave function effect, except for a slight
change in the kinematics between the two reactions as explained in
Sect.\ 2.1. The solid curves are obtained using the pair wave
functions based on the correlation function of Ref. \cite{friar}.
These are presumably the most realistic predictions in this work. For
comparison, however, the parametrization of Ref. \cite{hajduk} is
used to calculate the dashed curves. It is of interest to note that
the wave function differences have a larger effect outside the
$\Delta$ region. This was also the case in positive pion absorption
on the quasideuteron \cite{pihe}. There the spin observables had a
clear trend in the energy dependence and the quasideuteron results
crossed the free reaction results at the $\Delta$-peak energy. At 220
MeV only the final state proton polarization is similar for the
different initial state wave functions.  To check the trends
with increasing energy, we performed also the calculation fairly well
above the $\Delta$ region at 420 MeV photon energy not shown here.
The decrease of the cross section continued and  also other variables
continued slowly the trends of Fig.\ 6. Only in the case of $A_y$ the
rate of change with energy between the different models is
significant at and above the $\Delta$ region. It seems that photon
energies below the $\Delta$ region are the most promising  to see
differences arising from the different initial states in absorption
on neutron-proton pairs.

To better understand the energy dependence of the observables we
present in Fig.\ 7 the magnitudes of the leading multipole
contributions to the total cross section as a function of energy
compared with those of the free deuteron photodisintegration (dashed
curves). Also in the quasideuteron case absorption above 200 MeV is
dominated by the direct isobar contribution Fig.\ 2d in the magnetic
dipole transitions, while below 200 MeV the electric dipole takes
over. In the free reaction the $\Delta$ becomes already dominant at a
somewhat lower energy resulting in a slightly deeper minimum in the
cross section. It can further be seen that the ratios of the
multipole strengths for the  quasifree and free cases remain quite
well energy independent above 100 MeV so that one would, indeed,
expect a rather smooth energy dependence. The strongest energy
dependence appears in the ratio M2$(qd)/$M2$(d)$. This multipole is
strongly affected by the $N\Delta$ admixture of the $^3F_3$ final
state. Also the strengths of all multipoles increase in going from
the free to the quasifree reaction so that the qualitative similarity
of the observables is understandable as far as the quasideuteron
absorption is concerned. However, because of the different quantum
numbers, obviously there must be more changes in the absorption on
the $^1S_0$ pairs. As discussed in Sect.\ 2.3, in that case the
$\Delta$ isobar cannot contribute directly, and in the corresponding
curves of Fig.\ 7b its explicit contribution to the amplitudes has
been left out also in the ``free deuteron'' comparison. However, the
$N\!\Delta$ component is retained in the calculation of the wave
functions. Without the isobar contribution the M1 multipole is
drastically suppressed and E1 is far more prominent than in the
quasideuteron and dominates the process now up to 300 MeV. The
``free'' reaction without the $\Delta$ would be completely dominated
by E1. Also the energy dependence of the relative multipole strengths
is now quite different,  apparently causing the stronger energy
dependence observed above in the analyzing power $A_y$, where the
$^1S_0$ pair amplitude appears in first order.

It is also of interest to point out a prominent feature in the E2
multipole transition in the $^1S_0$ pair case. The dip around 300 MeV
is due to the renormalizing effect of the strong $^5S_2(N\!\Delta)$
component on the $^1D_2(np)$ final state wave function, i.e., part of
the two-baryon wave function, i.e., the  $N\!\Delta$ component, is
not directly contributing to the reaction. The dotted curve shows the
E2 multipole contribution calculation using the two-nucleon wave
function without the coupling to the isobar configurations.  To our
knowledge such a strong feedback from an isobar admixture on the {\it
nucleonic} part of the wave function has not been observed elsewhere.
The unique selectivity of the isospin structure is responsible for
the large reduction by about a factor of two in the E2 contribution
to the total cross section.  It remains to check whether it can be
observed in two-body photoabsorption on the $pp$ pair in $^3$He,
where the E2 multipole should be far more important \cite{future}.

Finally, Fig.\ 8 shows a comparison to the recent experimental
results of the TAGX collaboration \cite{tagx}. The magnitude of the
total cross section is quite reasonable for both models of the pair
wave functions. For comparison, the dotted curve shows the free
deuteron result without the factor 1/2 used in Fig.\ 6. The energy
dependence below 200$\,$MeV is not well reproduced in any of the
models. It is possible that at low momentum transfers also the
spectator with its Fermi motion can contribute. Also in the data the
$\Delta$ peak is somewhat shifted towards lower energies as compared
to the free deuteron case and the present $^3$He calculations. The
earlier data both in this energy region \cite{dhose} and below
\cite{fetisov} would allow a somewhat higher cross section than Ref.\
\cite{tagx}.

\section{Summary and Conclusions}

We have studied photodisintegration of $^3$He in a quasideuteron
model where the photon is absorbed at a correlated $np$ pair while
the third nucleon merely acts as a spectator. In addition to a bound
quasideuteron, we have also considered the contribution of a bound
$^1S_0(np)$ pair. The final $np$ pair interaction is completely
included as well as MEC and $\Delta$ excitation. For the initial pair
correlation function several models have been used. The compression
of the $np$ correlation wave function compared to the free deuteron
case enhances both short-range MEC and $\Delta$ effects and the
one-body contribution as well. This nearly triples the cross section.
However, the quasideuteron cross section becomes reduced by a
statistical factor 1/2 leading to an overall enhancement by about 1.5
as compared with the free reaction. Also in the relative spin
observables and angular distributions there are significant changes
both well below and above the $\Delta$-resonance region. Around the
resonance, the changes in the nucleonic and meson currents on one
hand and in the isobar current on the other hand act in opposite
directions and thus cancel each other to some extent in the relative
quantities so that the $^3$He results are rather similar to the free
deuteron case. This insensitivity on the initial pair wave function
suggests that in experiments trying to use two-body reactions to
investigate these wave functions, the $\Delta$ region is not a good
choice. The same conclusion has also been reached in corresponding
pion absorption studies \cite{pihe} where a systematic energy
dependence of the change was seen. The sensitivity is far greater
outside this energy region, and significant effects can be expected
there.

In the present calculation the quasideuteron disintegration was by
far the dominant contribution to the observables up to the $\Delta$
region, while the $^1S_0$ initial configuration appeared to be
significant only through the first order interference term in the
analyzing power $A_y$. However, well above this energy the $\Delta$
isobar loses its prominence and also the purely nucleonic current
from the $^1S_0$ pair gains importance. So higher energies are more
sensitive to this component. There one could also expect a rising
importance of the excitation of the Roper resonance in the magnetic
dipole transitions to the $^3S_1 - ^3\! D_1$ waves.

For the $np$ final states both the triplet and singlet initial
configurations add coherently. However, the singlet pair could be
separately studied in diproton photodisintegration by mainly the $E2$
multipole transition \cite{future}. Any of these two-body break-up
studies may be within present experimental possibilities, e.g., LEGS
\cite{legs} or TAGX \cite{tagx}. One advantage of these two-body
photoabsorption processes is that just one or two quantum states can
be singled out in the initial states. It would also be interesting if
bremsstrahlung experiments could be performed with the relative final
state nucleon energy constrained so low that only the $S$-wave would
be sufficient for its description similarly to the present pion
production experiments \cite{e460}. In this kind of simplified
situation one could expect clean signals of different multipoles and
exchange currents to be seen and distinguished.\\[.5cm]

\section*{Acknowledgements}

J.A.N. thanks the Institut f\"ur Kernphysik of the University of
Mainz for hospitality during a visit when this work was started.
P.W.  acknowledges the hospitality of the Research Institute for
Theoretical Physics of the University of Helsinki.

\newpage
\pagestyle{empty}
\noindent {\bf FIGURE CAPTIONS}
\renewcommand{\labelenumi}{Fig.\arabic{enumi}}
\begin{enumerate}
\item \label{fig1}
{Comparison of the different pair wave functions used in the present
work: wave functions based on the correlation function of Friar et
al.~\protect{\cite{friar}} (dashed), on the Faddeev wave function
parametrization of Hajduk et al.~\protect{\cite{hajduk}} (dotted and
dash-dotted in case of the $^1S_0$ pair), and on the Bonn OBEPR
potential \protect{\cite{bonn}} (full).}

\item \label{fig2}
{Diagrammatic  representation of the various photoabsorption
mechanisms included in the present model.}

\item \label{fig3}
{Separate electromagnetic contributions to the cross section, photon
asymmetry $\Sigma$, proton polarization $P_y$, and analyzing power
$A_y$ for $\gamma+^{3}$He$\rightarrow (pn)+p_{spec}$ at
$k_{lab}=300\,$MeV: nucleonic one-body currents (dotted), normal part
(short-dashed), and consecutively added explicit $\pi/\rho\,$-MEC
(long-dashed) and $\Delta$ excitation (full).}

\item \label{fig4}
{As Fig.\ 3, but for the free deuteron photodisintegration reaction
$\gamma+d\rightarrow p+n$.}

\item \label{fig5}
{Dependence of various observables
for $\gamma+^3$He$\rightarrow (pn)+p_{spec}$ at $k_{lab}=300\,$MeV
on the different components of the initial pair wave functions:
only quasideuteron (short-dashed), only quasideuteron but without
$D$-state and with renormalized $S$-state (long-dashed),
only $^1S_0(np)$ pair (dotted), and complete
calculation (full). All components are based on the correlation
function of Ref.~\protect{\cite{friar}}.}

\item \label{fig6}
{Dependence of various observables for $\gamma+^3\mbox{He}\rightarrow
(pn)+p_{spec}$ at $k_{lab}=220$--$360\,$MeV on the model for the
initial pair wave functions: the wave function based on the
correlation function of Ref.~\protect{\cite{friar}} (full), on the
Faddeev wave function parametrization of Ref.~\protect{\cite{hajduk}}
(dashed), and the free deuteron OBEPR wave function with the cross
section multiplied by a factor $\frac{1}{2}$ (dotted).}

\item \label{fig7}
{The full curves show the leading multipole contributions to the
total photodisintegration cross sections for the quasideuteron
$\sigma_d$ (a) and the singlet $np$ pair $\sigma_s$ (b) as in Eq.\
(\ref{eq:he}) (without the factors 1/2 and 1/6) in comparison with
the free deuteron photodisintegration (dashed curves). The dotted
curve in (b) is explained in the text. In (b) the ``deuteron''
results contain only the  normal part.}

\item \label{fig8}
{The total cross section for two-body photoabsorption on $np$ pairs
in $^3$He. The curves as in Fig.\ 6 except that the deuteron cross
section is not multiplied by 1/2. The data are from Ref.\ \cite{tagx}.}
\end{enumerate}
\end{document}